\documentclass[journal]{IEEEtran}
\usepackage[utf8]{inputenc}
\usepackage{comment}
\usepackage{cite}
\usepackage{graphicx}
\usepackage{longtable}
\usepackage[utf8]{inputenc}
\usepackage{subcaption}
\usepackage{amsmath}
\usepackage{blindtext}
\usepackage{mathtools}
\usepackage{amssymb}
\usepackage{gensymb}
\usepackage{color}
\usepackage{authblk}
\usepackage{filecontents}
\usepackage{multirow}

\usepackage[table]{xcolor}
\usepackage{color}
\usepackage{adjustbox}

\begin{document}
\title{A Learning Framework for An Accurate Prediction of Rainfall Rates} 

\author{Hamidreza Ghasemi Damavandi}
\author{Reepal Shah}
\affil{Future H2O, Office of Knowledge Enterprise Development, Arizona State University, Tempe, AZ, US}
\date{}

\maketitle

\section{Abstract}

The present work is aimed to examine the potential of advanced machine learning strategies to predict the monthly rainfall (precipitation) for the Indus Basin, using climatological variables such as air temperature, geo-potential height, relative humidity and elevation. In this work, the focus is on thirteen geographical locations, called index points, within the basin. Arguably, not all of the hydrological components are relevant to the precipitation rate, and therefore, need to be filtered out, leading to a lower-dimensional feature space. Towards this goal, we adopted the gradient boosting method to extract the most contributive features for precipitation rate prediction. Five state-of-the-art machine learning methods have then been trained where pearson correlation coefficient and mean absolute error have been reported as the prediction performance criteria. The Random Forest regression model outperformed the other regression models achieving the maximum pearson correlation coefficient and minimum mean absolute error for most of the index points. Our results suggest the relative humidity (for pressure levels of 300 mb and 150 mb, respectively), the u-direction wind (for pressure level of 700 mb), air temperature (for pressure levels of 150 mb and 10 mb, respectively) as the top five influencing features for accurate forecasting the precipitation rate.


\section*{\textit{Keywords:}}
Artificial Neural Networks (ANNs), Gradient Boosting feature extraction, Indus Basin, Precipitation rate, Random Forest

\section{Introduction}
\begin{figure*}[!h]
    \centering
    \includegraphics{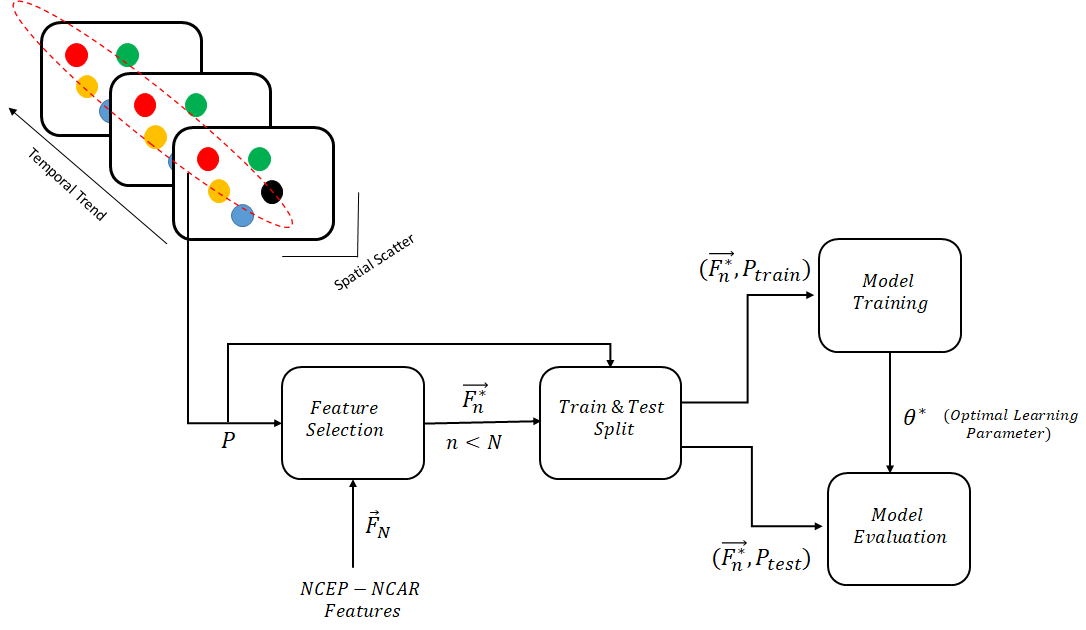}
    \caption{The block diagram of the precipitation rate prediction procedure via NCEP-NCAR features.}
    \label{bd}
\end{figure*}

Precipitation rate is a crucial concern for food production plan and water resource management. The prolonged dry period or heavy rain at the critical stages of the growth may lead to significant reduce crop yield. Additionally, accurate precipitation prediction could serve to alert the early warnings of sever weather events. Numerical prediction of precipitation rate has long been beneficial but a complex task for meteorologists, as it depends on other factors such as surface temperature (ST), humidity which often fluctuate in time and space. Hence, developing rigorous mathematical methods to accurately forecast the precipitation rate is of paramount importance. The motive is to harness the methodologies offered in data mining and machine learning fields to explore the other hydrological components in order to devulge valid and potentially useful internal relationships of these components as probable predictors to forecast the precipitation rate. Establishment of a machine learning model for precipitation prediction is rife with statistical assumptions to fine-tune the model’s hyper-parameters, and hence, reduce the computational time complexity and boost the prediction accuracy. Therefore, the artificial intelligence strategies can be considered as a highly efficient replacement for the traditional physical-based hydrological models. Several studies have focused on the implementation of machine learning techniques in precipitation prediction. Sumi et. al. 2012 \cite{sumi} investigated the potential of the artificial
neural network, multivariate adaptive regression splines, the k-nearest neighbour, and radial basis support vector regression for daily and monthly rainfall prediction of Fukuoka city in Japan. Hybrid models such as the combination of wavelet transform and artificial neural network (WANN) has been proposed by Kim et.al 2003 \cite{kim} where predictive models to predict the Conchos River Basin were proposed and evaluated. Karran et .al 2014 \cite{karran} compared the use of four different models, i.e. ANN, SVR, waveletANN, and wavelet-SVR in a Mediterranean, Oceanic, and Hemiboreal watershed. Khan et.al 2006 \cite{khan} examined the potential of Support Vector Regression (SVR) and Multilayer Linear Perception (MLP)in predicting lake water levels. Specifically, water level data for Lake Erie from 1918 to 2001 were used in training the two models to predict this property for the following 12 months; although the evaluation results (using root-mean-square and correlation coefficient) were promising in both cases, the intra-model comparison showed that the MLP outperformed SVR. Figure \ref{bd} illustrates the process to forecast the precipitation rate for the region of interest. We work out our precipitation on a index point by index point basis. For each index point, the precipitation rate is passed through the "Feature Selection" module where the most contributive predictors are determined. We withhold $90\%$ of the data for training the learning model and use the remaining $10\%$ for the evaluation. The optimal hyper-parameters are learned in the "Model Training" module and the performance of the model is evaluated by pearson correlation coefficient and mean absolute prediction error for the testing data in the "Model Evaluation" module.

The rest of the paper is organized as follows. Section \ref{Region_of_Interest} explains the region of the study and the dataset used in this work. The problem formulation and the methodology for extracting the informative features fore precipitation prediction are presented in Section \ref{methodology}. The most contibutive features as well as the performance of each learning model is illustrated in the Section \ref{result}. Section \ref{conclusion} concludes the paper.

We used precipitation data from 0.05 degree (5.528 Km) Climate Hazards Group InfraRed Precipitation with Station data (CHIRPS) v2 product. This product was developed by Funk et al. 2015 \cite{funk} by merging station-based data with satellite-based data. Initially global climatological mean was developed using station-based data from Food and Agricultural Organization (FAO) and Global Historical Climate Network (GHCN). This in total accounts for the total observations of $\sim 50,000$ stations across globe. Long-range precipitation derivatives were derived from Thermal Infra Red (TRI) Cold Cloud Duration (CCD) observation. Based on this two dataset, precipitation estimates were derived, named as CHRIP. Further, they merge station-based data from 2,00,000 locations from five different sources, generally observations decline from 32000 in 1980 to less than 14000 in 2014. This product is called CHIRPS. We took air temperature (Air), Geopotential height (Hgt), Relative humidity (Rhum), Specific humidity (Shum), Sea level pressure (Slp), u and v- direction wind as the potential predictors of the precipitation. The predictors were taken from NCEP-NCAR Reanalysis from 1981 to 2017 (37 years). Atmospheric variables in NCEP-NCAR reanalysis are simulated by assimilating station-based observation. Variables from NCEP-NCAR reanalysis were available at 2.5 degree. Hence precipitation from CHIRPS was aggregated for 2.5 degree resolution. Note that, these predictors were capture for different pressure levels (up to seventeen pressure levels) indicated in Table \ref{feature}. To be clear, two measured values for relative humidity at two different pressure levels are considered two unique potential predictors. Our constructed database would then consists of eighty-five unique predictors and their corresponding precipitation rate. These seventeen pressure levels (milli-bars) are 1000, 925, 850, 700, 600, 500, 400, 300, 250, 200, 150, 100, 70, 50, 30, 20 and 10. For simplicity, throughout this paper, we show these levels by $l_1$ to $l_{17}$. With this in mind, air temperature measured at first layer (1000 mb), would be represented as $ar^{l_{1}}$.

\section{Region of Interest} \label{Region_of_Interest}

The Indus basin with the total area of 1.12 million $km^2$ (432434.4176 $mi^2$) touches 4 countries of China (8 percent), India (39 percent), Afghanistan (6 percent) and Pakistan (47 percent). The Indus river originates from Lake Ngangla Ring Tsho in Tibetan plateau in the north, extending south to the dry plains in Pakistan and finally flows into the Arabian Sea. The rainfall in this basin is negligible and the snow and glaciers of the Hindu Kush - Himalayas and seasonal monsoon rains (July to September) are considered the main water input to this basin.
Being considered as the main source of water for agriculture, industrial productions, and human consumption, Indus basin Irrigates more than 30 million hectares of agricultural land in the basin through 15 tributaries including the Ravi, Beas, and Sutlej rivers in India, the Swat, Chitral,and Chenab rivers in Pakistan and Kabul River in Afghanistan. Hence, Indus river is considered as the vital water resource of the region.


\begin{figure*}[!h]
    \centering
    \includegraphics[scale=0.55]{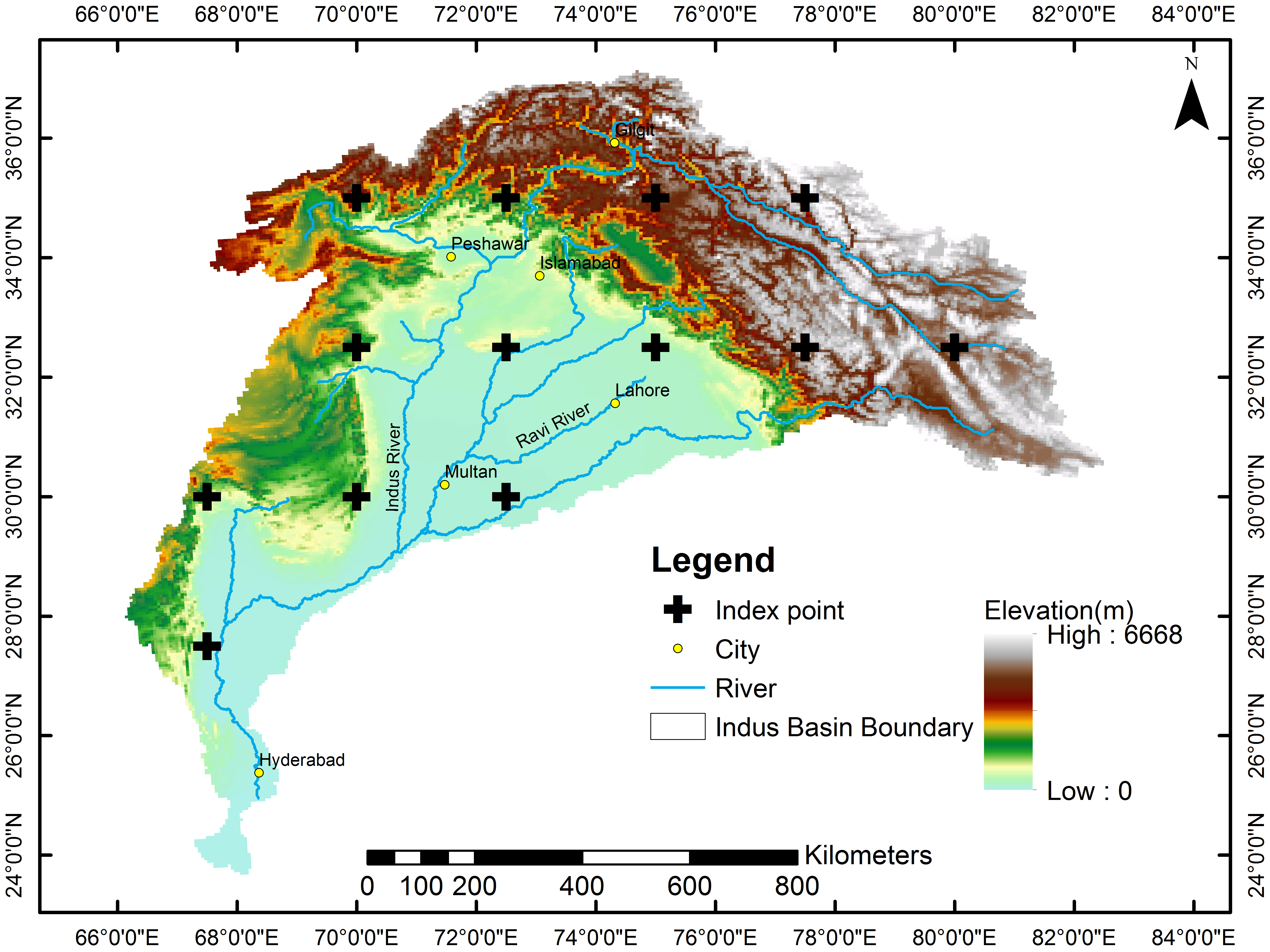}
    \caption{Indus basin: Region of Interest}
    \label{ROI}
\end{figure*}

\subsection{Dataset}

We used precipitation data from 0.05 degree (5.528 Km) Climate Hazards Group InfraRed Precipitation with Station data (CHIRPS) v2 product. This product was developed by Funk et al. 2015 \cite{funk} by merging station-based data with satellite-based data. Initially global climatological mean was developed using station-based data from Food and Agricultural Organization (FAO) and Global Historical Climate Network (GHCN). This in total accounts for the total observations of $\sim 50,000$ stations across globe. Long-range precipitation derivatives were derived from Thermal Infra Red (TRI) Cold Cloud Duration (CCD) observation. Based on this two dataset, precipitation estimates were derived, named as CHRIP. Further, they merge station-based data from 2,00,000 locations from five different sources, generally observations decline from 32000 in 1980 to less than 14000 in 2014. This product is called CHIRPS. We took air temperature (Air), Geopotential height (Hgt), Relative humidity (Rhum), Specific humidity (Shum), Sea level pressure (Slp), u and v- direction wind as the potential predictors of the precipitation. The predictors were taken from NCEP-NCAR Reanalysis from 1981 to 2017 (37 years). Atmospheric variables in NCEP-NCAR reanalysis are simulated by assimilating station-based observation. Variables from NCEP-NCAR reanalysis were available at 2.5 degree. Hence precipitation from CHIRPS was aggregated for 2.5 degree resolution. Note that, these predictors were capture for different pressure levels (up to seventeen pressure levels) indicated in Table \ref{feature}. To be clear, two measured values for relative humidity at two different pressure levels are considered two unique potential predictors. Our constructed database would then consists of eighty-five unique predictors and their corresponding precipitation rate. These seventeen pressure levels (milli-bars) are 1000, 925, 850, 700, 600, 500, 400, 300, 250, 200, 150, 100, 70, 50, 30, 20 and 10. For simplicity, throughout this paper, we show these levels by $l_1$ to $l_{17}$. With this in mind, air temperature measured at first layer (1000 mb), would be represented as $ar^{l_{1}}$.

\section{Methodology} \label{methodology}
\subsection{Problem Statement}
We attempt to predict the precipitation rate at an index point $\vec{l} = (lon,lat,elev)$ at time $t$ via other hydrological components such as air temperature, relative humidity .etc. It's noteworthy that, each layer within a hydrological component would also be treated as a potential predictor. In this work, we try to extract a subset of relevant features which can efficiently describe the input data
while reducing effects from noise or irrelevant variables and still provide promising prediction results. Specifically, we focus on gradient boosting method, expatiated in Section \ref{gradient_boosting}. Technically, we attempt to derive a function like $G(\cdot)$ to map the relevant observed variables (denoted in Table \ref{feature}) for a specific geographical location $\vec{l}$ and  sometime $t$ into a precipitation rate, deviated by some noise level of $w(\cdot)$.

\begin{equation}
    P(\vec{l},t) = G\big(\vec{F}(\vec{l},t)\big) + w(\vec{l,t})
\end{equation}

Towards this goal, five state-of-the-art learning models are trained on a location by location basis.

\subsection{Feature Selection}
Our dataset consists of $M$ examples (the monthly reported components of index points for Indus basin across the thirty-seven years) and $N$ potential predictors to forecast the precipitation rate ($M$ and $N$ equal to 5772 and 85, respectively). We focus on extracting the top $\kappa$ (here $\kappa$ is ten) most contributive predictors to forecast the precipitation rate. 

\subsubsection{Removing Linearly Correlated Features}

For any two potential predictors with the size of the examplse i.e. $M$, namely $f^M_i$ and $f^M_j$, we examine their linear correlation via the cosine angle between them in the feature space:

\begin{equation}
    cos(\phi_{i-j}) = \frac{\langle f_i, f_j\rangle}{|f_i|_1.|f_j|_1}
\end{equation}

where $\langle \cdot \rangle$ and $|\cdot|_1$ determine the inner product and the $L_1$ norm, respectively. We call two predictors, "co-linear" variables, if the above-mentioned $|cos(\cdot)|$ is at least $\gamma$ (here $\gamma=0.9$). In the first phase of feature filtering procedure, we remove the highly correlated predictors. 

\subsubsection{Selecting Optimal Features of Predictive Model} \label{gradient_boosting}

As the next stage to filter out the irrelevant features, we exploit the gradient boosting method. Gradient boosting is a machine learning technique for regression problem, which produces a prediction model in the form of an ensemble of weak prediction models, typically decision trees. Given the dataset of size $K$, $\{y_i, x_i \}^{i=K}_{i=1}$, we attempt to fit a model $F_m(x)$ to minimize the square loss at the $m^{th}$ stage of the algorithm. Such imperfect model, or the so-called weak model, would lead into some non-zero regression error. We now add an additional model $h$ to $F$ towards reducing this regression error, and hence the new model would be $F_{m+1}(x) = F_m(x) + h(x)$. Such new model would ideally be chosen to lead into a zero regression error, or $F_{m+1}(x_i) = F_m(x_i) + h(x_i) = y_i \forall 1 \le i \le K$ or equivalently $h(x_i) = y_i - F_m(x_i)$.
In other words, in each stage, we add a new model $h(\cdot)$ to the existing model to compensate its shortcomings, or the regression error. $y_i - F_m(x_i)$ $\forall 1 \le i \le K$ are called the residuals. Now towards identifying such $h(\cdot)$ model, we would fit a curve to the residuals, i.e. fit a curve to the data set $\{y_i - F_m(x); x_i \}^{i=K}_{i=1}$. Such fitted curve would then be added to construct $F_{m+1}(x)$. Obviously this new model is a stronger model with respect to $F_m(x)$ leading to a lower regression error. In case the regression error with the new model $(F_{m+1}(x))$ is still unsatisfactory, we would construct $F_{m+2}(x)$ via adding a new $h(x)$ to $F_{m+1}(x)$ with the same previous procedure. 

We now show that this idea is equivalent to updating the model using the gradient descent. Consider the dataset $\{y_i); F(x_i)\}^{i=K}_{i=1}$ with an attempt to minimize the square loss of $J = L(y,F(x)) = \sum_{i=1}^K \big[\frac{1}{2} (y_i-F(x_i))^2\big]$. The gradient descent of $y$ with respect to $F(x)$ is:

\begin{equation}
\frac{\partial J}{\partial F(x_i)} = \frac{\partial \sum_{i=1}^K \big[\frac{1}{2} (y_i-F(x_i))^2\big]}{\partial F(x_i)} = F(x_i) - y_i = - h(x_i) 
\end{equation}

and the model at $(m+1)^{th}$ stage is:

\begin{equation} \label{model_gd}
    F_{m+1}(x_i) = F_m(x_i) - \rho \nabla J_{F_m(x_i)} = F_m(x_i) + h(x_i)
\end{equation}
\begin{equation*}
    \forall ~ 1 \le i \le K ~,~ \rho = 1
\end{equation*}
We fine-tune the trained model $F(x)$ using Equation \ref{model_gd} until the regression error is desirably low. Note that at each stage $m$, we construct 100 regression trees and for each input $x_i$, $F(x_i)$ would be the average regressed values of all of these trees. It is noteworthy to mention that each tree would adopt a subset of the features to regress the input. We rank the importance of the features in terms of the \textit{number of occurrence} in each of the regression trees. Hence, features will be sorted in terms of contribution to reach the minimum regression error. We acquire the top 10 contributive features and pass on to the learning phase.

\begin{table}[!h] 
\centering

 \begin{tabular}{||c||c || c||} 

 \hline
 Index & Feature Name & Pressure Levels (Millibars)\\ [0.5ex] 
 \hline\hline
 1-17 & Air for 17 pressure layers & $l_1$ \textit{to}~ $l_{17}$ \\
 \hline
 18-34 & Hgt for 17 pressure layers & $l_1$ \textit{to}~ $l_{17}$\\
 \hline
 35-42 & Rhum for 8 pressure levels & $l_1$ \textit{to}~ $l_{8}$\\
 \hline
 43-50 & Shum for 8 pressure levels & $l_1$ \textit{to} ~ $l_{8}$\\
 \hline 
  51 & Slp for one pressure level & $l_1$\\ [1ex] 
  \hline
 
  52-68 & uwnd for 17 pressure levels & $l_1$ \textit{to}~ $l_{17}$\\ [1ex] 
  \hline

 69-85 & vwnd for 17 pressure levels & $l_1$ \textit{to}~ $l_{17}$\\ [1ex] 
  \hline

 \hline
\end{tabular}

\caption{Hydrological features used to predict the precipitation rate.}
\label{feature}
\end{table}

\subsection{Models}

Five state-of-the-art machine learning models, coupled by the gradient boosting feature selection technique, have been trained and their performance have been reported individually. We examined linear regression, K-nearest neighborhood regression $(K=3)$, random forest, support vector regression and multilayer perception as the predictive model.

\section{Result} \label{result}

Let $P(\vec{l},t)$ and $\tilde{P}(\vec{l},t)$ be the actual and the predicted precipitation rate at a particular geographical location $(lon,lat,elev)$ at sometime $t$. We evaluate the performance of the predictive learning model via the pearson correlation coefficient $(\rho)$ (Equation \ref{rho_def}) between the actual and predicted rates, mean absolute prediction error $(\mu)$ (Equation \ref{mu_def}) and its standard deviation $(\sigma)$. Table \ref{comparison_table} represents these parameters for each of the geographical location examined in this work for five different learning models, i.e., Random Forests (RF), $K$-nearest Neighborhood Regression ($K$-NN), Support Vector Regression (SVR), Linear Regression (LR), Multi-linear Perceptron (MLP). It is noteworthy to mention that for each geographical location we have used $90\%$ of the 37 years data and have evaluated the learning model against the remaining $10\%$ datapoints. \begin{equation}\label{rho_def}
\rho(\vec{l}) =  \frac{cov[P(\vec{l}),\tilde{P}(\vec{l})]}{\sigma_P . \sigma_{\tilde{P}}}\\ , ~~~
|\rho(\cdot)| \le 1
\end{equation}

\begin{equation}\label{mu_def}
\mu(\vec{l}) = \frac{\sum_{t=1}^N|P(\vec{l},t) - \tilde{P}(\vec{l},t)|_1}{N}
\end{equation}

Where $|\cdot|$ represents the absolute value.
Table \ref{comparison_table} summarizes the correlation between the predicted and observed values as well as the mean absolute prediction error. As shown, the correlation coefficient ranges from 0.53 in the last index point to 0.93 in the six-th index point. As can be seen here, Random Forest model outperformed the other models, nominated as best predictive model for six of the index points. Note that, these predicted values where calculated just by the internal relation among the hydrological models, computed in order of seconds, but could achieve the correlation coefficient of up to 0.93. This is an indicative of the potential of the learning models to emulate the complex equations governing the relationship between these components. Table \ref{optimal_feature} illustrates the top ten contributive features for each of the geographycal region. Table \ref{feautre_occurr} represents the number of occurrence of each of the features as the top ten contributive features. According to this table, $rh^{l_1}$ turned out to be the most frequent feature followed by $rh^{l_8}$, $uw^{l_4}$, $ar^{l11}$ and $ar^{l_{17}}$ as the top five feature playing pivotal role in the precipitation prediction. Due to the strong correlation between humidity and precipitation rate, it is of no surprise to determine the relative humidity as the most contributive feature for precipitation rate. 
We infer this observation by the direct correlation between the  air temperature and humidity to the precipitation rate \cite{breth} as the humidity can be considered as both the precursor and the result of rain. Our results indicating the importance of wind speed for precipitation rate aligns with some the established prior work e.g. \cite{hsu}, \cite{back} and \cite{johansson}.

\begin{table*}[!t]
\centering
 \begin{tabular}{||c || c|| c || c|| c||}
\hline
 Coordinate (longitude, latitude, elevation)& Learning Model & Pearson Coeff. ($\rho$) & MAE ($\mu$) & STD ($\sigma$)\\ 
 \hline\hline
 \textbf{\multirow{-4}*{(27.5 ,  67.5 ,  472.9)}} &
 \shortstack{ \\ RF \\ K-NN \\ SVR \\ LR \\ \color{blue}{MLP}}&
 \shortstack{0.89 \\ 0.87 \\ 0.47 \\ 0.88 \\ \color{blue}{\textbf{0.91}}}&  \shortstack{3.59 \\ 3.74 \\ 5.34 \\ 3.85 \\ 3.7} &
 \shortstack{3.34 \\ 3.84 \\ 7.92 \\ 3.29 \\ 2.89}\\ 
 
 \hline
 
 \textbf{\multirow{-4}*{(30. ,   67.5 , 1232.5)}}&
 \shortstack{ \\ \color{blue}{RF} \\ K-NN \\ SVR \\ LR \\ MLP}   & \shortstack{\color{blue}{\textbf{0.8}} \\ 0.75 \\ 0.39 \\ 0.77 \\ 0.75}  &  \shortstack{6.06 \\ 6.25 \\ 9.09 \\ 6.56 \\ 7.01} &
 \shortstack{5.68 \\ 6.49 \\ 9.47 \\ 5.64 \\ 5.67}\\ 
 
 \hline
 
 \textbf{\multirow{-4}*{(30. ,  70. , 721.5)}} & 
 \shortstack{ \\ \color{blue}{RF} \\ K-NN \\ SVR \\ LR \\ MLP}   & 
 \shortstack{ \color{blue}{\textbf{0.88}} \\ 0.87 \\ 0.57 \\ 0.83 \\ 0.87}  &  \shortstack{8.46 \\ 6.73 \\ 12.11 \\ 8.96 \\ 8.15} &
 \shortstack{7.97 \\ 8.63 \\ 16.16 \\ 8.01 \\ 7.05}\\ 
 
 \hline

 \textbf{\multirow{-4}*{(30. ,  72.5 , 148.2)}} & 
 \shortstack{ \\ RF \\ K-NN \\ SVR \\ \color{blue}{LR} \\ MLP}   &
 \shortstack{ 0.88 \\ 0.76 \\ 0.46 \\ \color{blue}{\textbf{0.89}} \\ -0.64}  &  \shortstack{8 \\ 10.27 \\ 14.75 \\ 8.55 \\ 68.17} &
 \shortstack{7.86 \\ 10.66 \\ 16.95 \\ 6.15 \\ 24.49}\\ 
 
 \hline
 
\textbf{\multirow{-4}*{(32.5 ,   70. ,  1203.)}} &
\shortstack{ \\ \color{blue}{RF} \\ K-NN \\ SVR \\ LR \\ MLP}   &
\shortstack{ \color{blue}{\textbf{0.9}} \\ 0.73 \\ 0.7 \\ 0.8 \\ 0.73}  &  \shortstack{8.36 \\ 11.28 \\ 13.64 \\ 11.45 \\ 15.24} &
\shortstack{7.14 \\ 10.38 \\ 13.28 \\ 6.51 \\ 10.1}\\ 
 
 \hline
 
\textbf{\multirow{-4}*{(32.5 ,  72.5,  326.4)}} &
\shortstack{ \\ RF \\ \color{blue}{K-NN} \\ SVR \\ LR \\ MLP}   &
\shortstack{ 0.85 \\ \color{blue}{\textbf{0.93}} \\ 0.65 \\ 0.8 \\ 0.82}  &  \shortstack{18.81 \\ 12.85 \\ 27.17 \\ 21.52 \\ 20.75} &
\shortstack{18.34 \\ 12.62 \\ 34.21 \\ 19.05 \\ 17.51}\\ 
 
 \hline

\textbf{\multirow{-4}*{(32.5,  75.,  967.7)}} & 
\shortstack{ \\ \color{blue}{RF} \\ K-NN \\ SVR \\ LR \\ MLP}   & 
\shortstack{ \color{blue}{\textbf{0.79}} \\ 0.71 \\ 0.18 \\ 0.77 \\ 0.34}  &  \shortstack{34.89 \\ 36.83 \\ 49.02 \\ 40.25 \\ 73.55} &
\shortstack{44.33 \\ 51.79 \\ 80.93 \\ 42.91 \\ 89.66}\\ 
 
 \hline
 
\textbf{\multirow{-4}*{(32.5,   77.5, 4044.4)}} &
\shortstack{ \\RF \\ \color{blue}{K-NN} \\ SVR \\ LR \\ MLP}   &
\shortstack{ 0.63 \\ \color{blue}{\textbf{0.66}} \\ 0.29 \\ 0.6 \\ -0.27}  &  \shortstack{27.89 \\ 25.57 \\ 28.12 \\ 32.7 \\ 304.86} &
\shortstack{24.42 \\ 22.99 \\ 37.43 \\ 19.94 \\ 46.9}\\ 
 
 \hline
 
\textbf{\multirow{-4}*{(32.5,   80.,  4882.1)}} & 
\shortstack{ \\ \color{blue}{RF} \\ K-NN \\ SVR \\ LR \\ MLP} &
\shortstack{ 0.68 \\ 0.48 \\ \color{blue}{\textbf{0.68}} \\ 0.61 \\ 0.26}  &  \shortstack{11.23 \\ 13.36 \\ 12.74 \\ 12.23 \\ 120.48} &
 \shortstack{9.85\\ 14.17 \\ 14.73 \\ 10.13 \\ 19.38}\\ 
 
 \hline
 
\textbf{\multirow{-4}*{(35.,    70.,  2409.8)}} & 
\shortstack{ \\ RF \\ \color{blue}K-NN \\ SVR \\ LR \\ MLP}   & 
\shortstack{ 0.76 \\ \color{blue}{\textbf{0.83}} \\ 0.59 \\ 0.75 \\ 0.82}  &  \shortstack{12.77 \\ 12.89 \\ 18.95 \\ 13.14 \\ 11.55} &
 \shortstack{10.57 \\ 10.37 \\ 16.02 \\ 10.23 \\ 8.89}\\ 
 
 \hline

\textbf{\multirow{-4}*{(35,    72.5, 2256.2)}} & 
\shortstack{ \\ \color{blue}{RF} \\ K-NN \\ SVR \\ LR \\ MLP}   &
\shortstack{ \color{blue}{0.75} \\ 0.67 \\ 0.59 \\ 0.66 \\ 0.38}  & 
\shortstack{20.66\\ 21.63 \\ 26.32 \\ 23.63 \\ 33.94} &
\shortstack{16.92 \\ 18.54 \\ 20.95 \\ 17 \\ 17.87}\\ 
 
 \hline
 
\textbf{\multirow{-4}*{(35.,    75.,  3590.8)}} & 
\shortstack{ \\ \color{blue}{RF} \\ K-NN \\ SVR \\ LR \\ MLP}   & 
\shortstack{ \color{blue}{\textbf{0.55}} \\ 0.25 \\ 0.17 \\ 0.51 \\ 0.24}  &  
\shortstack{16.93 \\ 19.65 \\ 16.33 \\ 16.1 \\ 17.39} & 
\shortstack{10.13 \\ 14.31 \\ 13.81 \\ 12.05 \\ 12.45}\\
 \hline
 
 \textbf{\multirow{-4}*{(35.,    77.5, 4892.9)}} &
 \shortstack{ \\ RF \\ K-NN \\ SVR \\ \color{blue}{LR} \\ MLP}   &
 \shortstack{ 0.4 \\ 0.49 \\ 0.4 \\ \color{blue}{\textbf{0.53}} \\ 0.09}  &  \shortstack{9.27 \\ 8.68 \\ 8.32\\ 8.88 \\ 77.02} &
 \shortstack{6.8 \\ 6.61 \\ 7.76 \\ 6.2\\ 12.12}\\ 
 
 \hline
 \end{tabular}
\caption{Comparison between different machine learning models in terms of pearson correlation coefficient, mean absolute error (MAE) and  standard Deviation of error of the predicted precipitation rate.}
\label{comparison_table}
\end{table*}

\begin{table*}[t]
\centering
\begin{tabular}{||c||c||} 
 \hline
 Coordinate (lon, lat, elev) of the index point & Top 10 Features \\ [1ex] 
 \hline\hline
 $(27.5 ,  67.5 ,  472.9)$ & $\textit{rh}^{l_1,l_3,l_5}, ar^{l_{12}}, vw^{l_4,l_5,l_{12}}, sh^{l_6,l_7}, uw^{l_4}$  \\  \hline
 
 $(30. ,   67.5 , 1232.5)$ & $\textit{rh}^{l_1,l_4,l_5,l_6}, ar^{l_{11}}, vw^{l_1,l_{12}}, sh^{l_7}, uw^{l_1,l_4}$  \\ \hline
 
 $(30. ,  70. , 721.5)$ & $\textit{rh}^{l_1,l_4,l_7}, vw^{l_4,l_{14}, l_{15},l_{16}}, sh^{l_1,l_7}, uw^{l_4}$  \\ \hline
 
 $(30. ,  72.5 , 148.2)$ & $\textit{rh}^{l_1,l_4}, vw^{l_{12},l_{16}}, sh^{l_6,l_7}, uw^{l_2,l_3,l_4}, hg^{l_5}$ \\ \hline
 
 $(32.5 ,   70. ,  1203.)$ & $\textit{rh}^{l_1,l_8}, ar^{l_{11}}, vw^{l_1,l_4,l_{15},l_{16},l_{17}}, sh^{l_1}, uw^{l_1}$\\  \hline
 
 $(32.5 ,  72.5,  326.4)$ & $ \textit{rh}^{l_1,l_2}, ar^{l_1,l_{13},l_{16}}, vw^{l_1,l_{15}},sh^{l_1},uw^{l_4,l_5}$ \\ \hline
 
 $(32.5,  75.,  967.7)$ & $\textit{rh}^{l_1,l_7,l_8}, ar^{l_{11},l_{17}}, vw^{l_7}, sh^{l_1}, uw^{l_4,l_5}, hg^{l_5}$ \\ \hline
 
 $(32.5,   77.5, 4044.4)$ & $\textit{rh}^{l_1,l_5,l_6,l_8}, ar^{l_{11},l_{17}}, vw^{l_6,l_7}, uw^{l_6}, hg^{l_5}$\\ \hline
 
 $(32.5,   80.,  4882.1)$ & $\textit{rh}^{l_1,l_7,l_8}, ar^{l_{13},l_{14},l_{17}},vw^{l_7,l_{13}},sh^{l_8},,hg^{l_5}$ \\ \hline
 
 $(35.,    70.,  2409.8)$ & $\textit{rh}^{l_1,l_4,l_5,l_8}, ar^{l_{11},l_{15}},vw^{l_1,l_{16}},uw^{l_1,l_4}$ \\ \hline
 
 $(35,    72.5, 2256.2)$ & $\textit{rh}^{l_8}, ar^{l_1,l_7},vw^{l_{13},l_{15},l_{17}},sh^{l_6},uw^{l_1,l_4,l_5}$\\ \hline
 
 $(35.,    75.,  3590.8)$ & $\textit{rh}^{l_1,l_5,l_8}, ar^{l_{11},l_{17}},vw^{l_7,l_{13},l_{16},l_{17}},sh^{l_7}$\\ \hline
 
 $(35.,    77.5, 4892.9)$ & $\textit{rh}^{l_8}, ar^{l_{11},l_{15},l_{16},l_{17}},vw^{l_6,l_7,l_{16}}, uw^{l_1},hg^{l_4}$\\ \hline
 \end{tabular}
\caption{Top ten contributive predictors for precipitation prediction.}
\label{optimal_feature}
\end{table*}

\begin{table*}[h!]
\centering
\begin{tabular}{||c || c ||} 
 \hline
 Feature Name & Frequency of Occurrence\\ [0.5ex] 
 \hline\hline
\rowcolor{green} $\textit{rh}^{l_1}$ & $11$ \\ 
\rowcolor{green} $\textit{rh}^{l_8}$ & $8$ \\
 \rowcolor{green}$\textit{uw}^{l_4}$ & $8$ \\

 \rowcolor{green}$\textit{ar}^{l_{11}}$ & $7$ \\ 
 \rowcolor{green}$\textit{ar}^{l_{17}}$ & $6$ \\
 
 \rowcolor{green}$\textit{vw}^{l_1}$ & 6 \\ 
 
 \rowcolor{green}$\textit{rh}^{l_{5}}$ & 5 \\
 
 \rowcolor{green}$\textit{sh}^{l_7}$ & 5 \\
 
 \rowcolor{green}$\textit{uw}^{l_1}$ & 5 \\
 
 \rowcolor{green}$\textit{vw}^{l_{7}}$ & 5 \\ 
 
 \rowcolor{green!50}$\textit{hg}^{l_{5}}$ & 4 \\ 
 
 \rowcolor{green!50}$\textit{rh}^{l_{4}}$ & 4 \\ 
 
 \rowcolor{green!50}$\textit{sh}^{l_1}$ & 4 \\
 
 \rowcolor{green!50}$\textit{vw}^{l_{1}}$ & 4 \\
 
 \rowcolor{green!50}$\textit{vw}^{l_{15}}$ & 4 \\ 
 
 \rowcolor{green!50}$\textit{rh}^{l_{7}}$ & 3 \\ 
 
\rowcolor{green!50} $\textit{sh}^{l_6}$ & 3 \\ 

 \rowcolor{green!50}$\textit{uw}^{l_{5}}$ & 3 \\ 
 
 \rowcolor{green!50}$\textit{vw}^{l_{4}}$ & 3 \\ 
 
 \rowcolor{green!50}$\textit{vw}^{l_{12}}$ & 3 \\ 
 
 \rowcolor{green!25}$\textit{vw}^{l_{13}}$ & 3 \\ 
 
 \rowcolor{green!25}$\textit{vw}^{l_{17}}$ & 3 \\ 

 \rowcolor{green!25}$\textit{ar}^{l_1}$ & 2 \\ 
 
 \rowcolor{green!25}$\textit{ar}^{l_{13}}$ & 2 \\ 
 \rowcolor{green!25}$\textit{ar}^{l_{15}}$ & 2 \\ 
\rowcolor{green!25} $\textit{ar}^{l_{16}}$ & 2 \\ 

 \rowcolor{green!25}$\textit{rh}^{l_{6}}$ & 2 \\ 

\rowcolor{green!25} $\textit{vw}^{l_6}$ & 1 \\ 

\rowcolor{green!25} $\textit{ar}^{l_{12}}$ & 1 \\ 
\rowcolor{green!25} $\textit{ar}^{l_{14}}$ & 1 \\ 

\rowcolor{green!10} $\textit{hg}^{l_4}$ & 1 \\

\rowcolor{green!10} $\textit{rh}^{l_2}$ & 1 \\

\rowcolor{green!10} $\textit{rh}^{l_3}$ & 1 \\

\rowcolor{green!10} $\textit{sh}^{l_8}$ & 1 \\

\rowcolor{green!10} $\textit{uw}^{l_2}$ & 1 \\ 

\rowcolor{green!10} $\textit{uw}^{l_3}$ & 1 \\

\rowcolor{green!10} $\textit{uw}^{l_6}$ & 1 \\ 

\rowcolor{green!10} $\textit{vw}^{l_5}$ & 1 \\ 

\rowcolor{green!10} $\textit{vw}^{l_6}$ & 14 \\ 
 \hline
\end{tabular}
\caption{The frequency of selection of each feature as the top ten contributive predictors towards precipitation prediction.}
\label{feautre_occurr}
\end{table*}


\section{conclusion} \label{conclusion}
Motivated by the recent advances in data-driven models across miscellaneous areas such as environmental and \cite{gh, gh1} and hydrological sciences \cite{sumi}, as a powerful tool to approximate the physical-based models, this study aimed to evaluate the performance of the artificial intelligence (AI) data-driven models, and in particular, the top state-of-the-art machine learning models to predict the precipitation rate using using other hydrological components listed in Table \ref{feature} over the region of interest i.e. Indus Basin. We reported the prediction performance as the pearson correlation coefficient and mean square error (MAE defined in Equation \ref{mu_def}) between the observed and predicted precipitation rates. Random forest model outperformed the others, achieving the maximum pearson coefficient and minimum mean absolute error for most of the index points. We used seven unique hydrological components shown in Table \ref{feature}, each measured in different pressure levels. We treated a component measured at particular pressure level as an individual feature leading to eighty-five distinct potential predictors to forecast the precipitation rate. We harnessed the gradient boosting strategy to extract the relevant features to an accurate prediction. The results reveal the importance of relative humidity, u-direction of wind and air temperature as the prominent predictors to estimate the precipitation rate. 

The promising results in this work certify the prominent opportunities provided by the artificial intelligence (AI) methods, and particularly machine learning models, to mine the underlying concealed knowledge 
between diverse hydrological components which is otherwise not readily accessible to the researchers. Although the conventional hydrological models have been well-established but suffer from some shortcomings such as the tedious computational time or the uncertainties in variables estimate. The AI methods have been proven to be effective in tackling these deficiencies while being backed by advanced theories in optimization, and hence, offering efficient alternative for the conventional physical-based hydrological models.

\section*{Acknowlegment}
This work was supported by the National Science Foundation award (grant: GR10458) and conducted at Future H2O, Office of Knowledge Enterprise Development (OKED) at Arizona State University. 

\bibliographystyle{unsrt}
\bibliography{refs}

\end{document}